# 1D and 2D optical lattices on a chip for quantum computing

Katharina Christandl and Gregory P. Lafyatis

Department of Physics, The Ohio State University, Columbus, Ohio 43210

Seung-Cheol Lee and Jin-Fa Lee

Department of Electrical Engineering, The Ohio State University, Columbus, Ohio 43210

We propose a way to make arrays of optical frequency dipole-force microtraps for cold atoms above a dielectric substrate. Traps are nodes in the evanescent wave fields above an optical waveguide resulting from interference of different waveguide modes. The traps have features sought in developing neutral atom based architectures for quantum computing: ~ 1 mW of laser power yields very tight traps 150 nm above a waveguide with trap vibrational frequencies ~ 1 MHz and vibrational ground state sizes ~ 10 nm. The arrays are scalable and allow addressing of individual sites for quantum logic operations.

Requirements for the physical implementation of quantum computing are a quantum two-level system (qubit) for storing quantum information, the ability to initialize and read out the qubits, a set of universal quantum gates (one- and two-qubit gates), long decoherence times, and scalability.[1] Recent cold neutral atom based quantum computing schemes [2,3] share several features. The qubit consists of two hyperfine states of an individual atom, tightly confined in a "microtrap." One-qubit operations may be effected using either microwave fields or optically, using stimulated Raman transitions. Reading out or projecting the state of a qubit would be accomplished using laser-induced fluorescence. The most challenging requirement for quantum computing is performing the two-qubit operations needed for a universal quantum computer. Two qubits must be selected and controllably made to interact and entangle. Schemes using the "dipolar" interaction [4] or "controlled collisions" [5] for entanglement have been investigated. To perform quantum computations of interest with error correction ~ $10^4$ qubits are required. Therefore, microtrap based schemes containing one atomic qubit per trap need to be scaled up to these numbers. One possible way of doing this is using optical lattices. An optical lattice is the intensity pattern formed when two or more beams from a laser are made to intersect. The nodes or antinodes of such a pattern can serve as an array of microtraps for ultracold atoms. So far as we know, *all* proposed large-scale implementations of quantum computing using 3D optical lattices have difficulties addressing individual atoms, either to select specific atoms for one or two qubit gate operations or for reading out. We have found a way of creating *1D and 2D optical lattices on a chip* and discuss how these may solve the qubit addressing difficulty and allow scaling up qubit registers.

The prospect of a wide variety of applications has helped spawn recent research efforts towards miniaturization of atom optics components on condensed matter substrates --- atom chips. Two major approaches to controlling atoms on atom chips are magnetic or opto-magnetic atom manipulation [6] and evanescent wave (*EW*) based techniques that exploit the "dipole-force." [7,8].

Optical dipole-force traps for atoms [9] are usually made by creating either a local maximum (for laser light red-detuned from an atomic resonance, $\omega_0$) or minimum (for blue-detuned light) of light intensity at an isolated point in space. By interfering laser beams in free space, one can create an optical lattice, a 3D array of sub-wavelength sized microtraps located at the nodes or antinodes of the interference pattern. The spacing between traps is $\lambda/2$, where $\lambda$ is the wavelength of the trapping lasers. Traps near condensed matter surfaces may be made using the evanescent optical frequency electric fields created by total internal reflection of light traveling within the condensed matter substrate, but an *outstanding problem is to arrange to have the desired intensity extrema isolated in space and not at the vacuum-surface interface*. Previously, this problem has been circumvented either by using two colors of laser light --- a red detuned *EW* that attracts atoms toward a surface and a second blue detuned wave keeping them from actually hitting the surface --- or by using additional forces for trapping such as gravity or the atom-surface van der Waals force [7].

Relevant previous theoretical work includes the proposal of Barnett et al. to use two-color *EWs* above a slab-type *optical* waveguide to create a *cold atom* waveguide.[8] Burke et al. have proposed several waveguide based *EW* traps and their "dark-spot" trap is the closest the previous literature comes to the present work.[7]

The present *experimental* state-of-the-art in *EW* traps is the two color trap of Hammes et al. [10] in which about 20000 Cs atoms were confined to a ~ 0.3 mm diameter pancake shaped region about 1 micrometer above a prism. Dumke et al.[11] have realized an array of microtraps by trapping $^{85}Rb$ atoms at the foci of a laser beam passed through a substrate patterned with an array of micro-lenses. Those traps were spaced 125 $\mu$ m apart and confined up to 1000 atoms each, 625 $\mu$ m above the substrate surface.

The essential new idea of our work is illustrated in Fig. 1(a). A planar optical waveguide consists of a guiding dielectric film with index of refraction $n_W$ over a dielectric substrate with index $n_S$, where $n_S < n_W$. We consider TE modes of light of wavelength $\lambda$ (frequency $\omega$) propagating along $z$, which have only an $x$-component $E_x$ of electric field and for which analytic solutions exist.[12] A traveling wave in the $m^{th}$ mode propagates with a spatial phase of $\exp[ik_m z]$ determined by $k_m = n_{eff,m} k$, where $k = 2\pi/\lambda$ and $n_{eff,m}$ is the "effective index of refraction" of mode $m$. It can be shown that $n_S < n_{eff,m} < n_W$ and $n_{eff,m}$ decreases with increasing mode order up to "cutoff" --- the highest order guided mode. The electric field of a mode decays in the evanescent region above the waveguide as $\sim \exp[-\delta_m y]$, where $\delta_m = \sqrt{k_m^2 - k^2}$. If two modes, "$a$" and "$b$," are fed the same frequency laser, the electric field in the vacuum above the waveguide is

$$E_x(y,z) = E_a \exp[-\delta_a y]\exp[ik_a z + i\theta_a] + E_b \exp[-\delta_b y]\exp[ik_b z + i\theta_b] \quad (1)$$

times a common time phase factor. $\theta_a$ and $\theta_b$ are phase constants, and $E_a$ and $E_b$ are the field amplitudes of the modes at the waveguide-vacuum interface, experimentally determined by the laser power in the respective modes. Along the waveguide there are values of $z$, occurring periodically with spatial period $\Delta z = \frac{2\pi}{|k_b - k_a|}$, where the two modes are exactly out of phase. By an appropriate choice of phases, we can write the field above the waveguide in one of these planes, $z^*$

$$E_x(y, z^*) = E_a \exp[-\delta_a y] - E_b \exp[-\delta_b y], \qquad (2)$$

and by adjusting the ratio of laser powers in the modes, we can create field nodes (nodal lines along $x$), at any desired height, $y^*$, above the waveguide (see Fig. 1(b)).

Now, a 2D optical lattice can be constructed by crossing two of these 1D waveguide structures, identical apart from their propagation directions, as shown in Fig. 1(c). Here, the waveguide confines light only in the $y$-direction. In the transverse direction along the waveguide, the input laser beams are collimated. In the area of overlap, a 2D array of nodal points is created, which, for laser light blue-detuned from resonance, can trap atoms. For the cases we consider, the potential energy of an atom in an optical frequency field is: [7]

$$U(\mathbf{r}) = \frac{2}{3} \frac{\hbar \Gamma}{8} \frac{\Gamma}{\Delta} \left| \frac{E(\mathbf{r})}{E_{sat}} \right|^2. \qquad (3)$$

Here, $\Gamma$ is the spontaneous emission decay constant, $\Delta$ is the detuning from resonance of the laser light and $E(\mathbf{r})/E_{sat}$ is the ratio of the local electric field to the on-resonance electric field that saturates the optical transition.

We modeled such a 2D *EW* optical lattice for trapping $^{85}$Rb atoms. In a quantum computer an $^{85}Rb$ atom could store a qubit using the $|F=2\rangle$ and $|F=3\rangle$ levels of the electronic ground state, $5s_{1/2}$. Laser light was tuned to the blue of the $5s_{1/2} \to 5p_{3/2}$ transition, $\lambda = 780$ nm. We used TE0 and TE1 modes of a planar waveguide that consisted of a 230 nm thick *As₂S₃* film, $n_W = 2.4$, on a *NaF* substrate, $n_S = 1.32$. The resulting 2D square array of nodes has a nearest neighbor spacing of 0.98 $\mu$m. To make strong traps with currently available materials, it is necessary to work *very* close to the waveguide surface. We designed traps to be located 150 nm above the waveguide surface. We arbitrarily picked laser powers so that the mode(s) requiring the most power, the TE0 mode(s), had 1 mW per $\mu$m of the width of the laser beam in the transverse direction. A power of 0.042 mW per $\mu$m of laser beam width in the TE1 mode(s) then resulted in nodes 150 nm above the surface. The laser tuning determines the depth of the microtraps and the spontaneous emission rate. A detuning of 1000 $\Gamma$ to the blue of resonance produces traps that are 1.5 mK deep. The microtraps are *extremely* tight: The vibrational frequencies for atoms oscillating in the trap are $f_x$, $f_z = 1.7$ MHz ($hf_z/k_B = 80$ $\mu$K), and $f_y = 2.5$ MHz ($hf_y/k_B = 120$ $\mu$K). The dimensions of the ground vibrational state of these microtraps, defined by $\beta_i \equiv \left(\frac{\hbar}{2\pi m_{Rb} f_i}\right)^{1/2}$, are $\beta_x$, $\beta_z$ = 8.4 nm and $\beta_y = 6.9$ nm.

The lattice as described is much closer to the surface than any previously demonstrated trap. Similar traps, 500 nm above the surface, would be only ~ 2 $\mu$K deep. At 150 nm, the atom-surface van der Waals potential energy $U_{VDW}$ is fairly small, $U_{VDW}/k_B \sim 5\mu K$. But for quantum information applications, an important question is whether the proximity of the dielectric surface will lead to decoherence. So far as we know, no one has yet observed or suggested any specific decoherence mechanisms --- the problem simply has not been studied. [13,14]

Initial loading of atoms into such an array of traps could be accomplished via a superfluid-Mott insulator transition analogous to the 3D optical lattice loading recently demonstrated by Bloch et al.[15] This transition automatically produces the desired one atom per site into the microtrap vibrational ground states. 2D Mott transitions *are* expected for boson species. Alternatively, 2D arrays could be loaded more conventionally: by "puddling" atoms in an evanescent wave above the waveguide and then simultaneously turning on the trapping fields and deep-cooling the atoms.[10]

For a dipole trap at a field node, in the two-level approximation, the rate of spontaneous emission events for an atom in the vibrational ground state, $\eta$, is related to the eigenfrequencies of the trap by

$$\eta = \frac{\pi}{2}(f_x + f_y + f_z)\frac{\Gamma}{\Delta}. \qquad (4)$$

For the array described above, this is 9.2 kHz. Spontaneous emission does not lead to the loss of trapped particles --- in contrast to magnetic trapping. It does, however, limit the coherence time of the qubit, which in turn limits the number of operations possible before decoherence occurs. This number should exceed ~ $10^4$ to enable quantum computation with error correction. Increasing the detuning decreases the spontaneous emission rate, but weakens the trap. In our example, increasing the detuning by a factor of 150 leads to 10 $\mu$K deep traps. Only the ground state in the weakened harmonic oscillator potential is bound and the spontaneous emission rate is reduced to <5 Hz. One can do better still by increasing the laser power and further increasing the laser detuning. Increasing the laser power by a factor of 10 is reasonable, increasing it by a factor of 1000 may be possible.

A translatable 1D lattice of atom traps, a shift register of qubits, may be constructed by destructively interfering modes of a channel or ridge waveguide. We found, numerically,[16] the 20 lowest order modes for a 0.45 $\mu$m thick by 1.0 $\mu$m wide $As_2S_3$ ridge-type guide on a *NaF* substrate, shown in Fig. 2. Modes are no longer pure TE or pure TM but we found that there were several pairs of modes that could interfere to make a periodic array of optical frequency field *EW* nodes, 3D atom traps, as depicted in Fig. 2. Confinement in the $y$ and $z$-directions is achieved as in the previous case by picking two modes with different waveguide propagation vectors, $k_a$ and $k_b$, and, in turn, with different *EW* decay lengths in the vacuum region. For the modes we chose, the $E_y$ component is largest and provides trapping in those dimensions. In addition, the chosen modes had components $E_x(\mathbf{r})$ and $E_z(\mathbf{r})$ that were anti-symmetric about $x = 0$, the center of the width dimension of the waveguide, and provided confinement in the $x$-direction. We chose the pair of modes that made the strongest traps. These had $k_a = 1.304 \times 10^7$ m$^{-1}$ and $k_b = 1.105 \times 10^7$ m$^{-1}$ and resulted in a linear array of nodes spaced 3.2 $\mu$m apart. This array can hold a register of qubits. As above, we picked the power of the stronger mode, $a$, to be 1 mW, and created nodes 150 nm above the waveguide surface with a power of 0.081 mW in mode $b$. A tuning of the laser field 1000 $\Gamma$ to the blue of resonance produces microtraps with a depth of 0.77 mK. $^{85}Rb$ atoms would be confined with vibrational frequencies $f_x$ = 1.8 MHz, $f_y$ = 1.9 MHz, and $f_z$ = 0.65 MHz, and ground state dimensions $\beta_x$ = 8.1 nm, $\beta_y$ = 7.7 nm, and $\beta_z$ = 13.3 nm. The spontaneous emission rate for an atom in the vibrational ground state of a microtrap is 6.8 kHz. As above, increasing the laser detuning trades trap strength for reduced spontaneous emission. The register of qubits may be shifted along the waveguide by adjusting the phases of the two modes.

One can imagine various quantum computing architectures using these 1D and/or 2D lattices. The surface geometry allows *selectively* driving one-qubit operations by stimulated Raman transitions and reading out individual qubit states using laser-induced fluorescence. The required laser beams may be introduced from above. So, reading the state of a qubit might be carried out with an apparatus similar to the read head of a CD player. In our 2D example the lattice periodicity ~ 0.98 $\mu$m. If more separation between qubits were required to allow selective addressing, it could be achieved by using a thicker waveguide. For example, a 2D lattice identical to that described above but using a 1000 nm thick waveguide would have nearest neighbor qubits separated by ~ 10 $\mu$m. Alternatively, for 1D lattices, laser beams could be introduced through additional waveguides orthogonal to the microtrap array.

One example of a two qubit gate using 1D registers works as follows. Two parallel 1D registers can be made within a few $\mu$ m of one another. Two designated qubits--- one in each register --- can be lined up adjacent to one another by adjusting the phases of the waveguide modes. By feeding the qubits an appropriate sequence of laser pulses the selected qubits can be excited into Rydberg states and entangled via electric dipole-dipole interaction. Note: in implementing this type of gate it is important to consider the possibility of decoherence, heating, or even atom-loss if an atom in a Rydberg state experiences a different optical potential than a ground state atom and the internal and external degrees of freedom of the atom/qubit become entangled. Safronova et al. [17] have suggested a solution: by carefully choosing the frequency of the lattice laser, the ac polarizabilities of ground state atoms and Rydberg state atoms can be made identical: i.e. all atoms see the *same* optical potential and the internal and external degrees of freedom remain unentangled. Their polarizability-matching schemes are readily applicable to the waveguide-based optical lattice qubit arrays we discuss in this paper.

We have described a way to create 2D and 1D periodic arrays of optical microtraps above optical waveguides. These are analogs of 3D optical lattices. For quantum computer applications, beneficial aspects of these microtrap arrays include: they require little power; the traps are extremely tight; all ground state hyperfine Zeeman states are confined; trapping at field nodes reduces spontaneous emission decoherence problems; the system is highly scalable, and, most importantly, individual qubits are readily addressable. The most problematic feature of these traps is that, to get strong trapping with waveguides made from existing materials and using low laser powers, it is necessary to work *very* close to the surface of the waveguide. At this time, the impact the surface will have on the coherence required for quantum information applications is an entirely open issue.

While we have focused on possible quantum information applications of the proposed optical lattices on chips, we conclude by noting other possible uses. The near-surface feature of the traps could be exploited to study atom-surface interactions. For example, atom-surface forces perturb the vibrational frequency of a trapped atom for the normal mode that is perpendicular to the surface and by accurately measuring this frequency as a function of distance above the surface, it should be possible to study those forces. We estimate that stable traps may be generated at distances from 10 nm to 500 nm above the waveguide surface. In this range, the character of the atom-surface force should change from a "van der Waals" force to "Casimir" force. More generally, regular optical lattices have long been used as model systems for diverse problems and recently have proved especially useful for studying fundamental solid state questions. We anticipate the trap arrays described above could be used in similar studies as model 1D and 2D periodic systems. [18]

# Figure Captions

**Fig. 1. (Color online) 2D optical lattice formed by interfering waveguide modes.**

**(a) Schematic diagram of TE0 and TE1 modes interfering in a planar waveguide. The resulting 2D traps confine atoms along $y$ and $z$.**

**(b) For laser powers of 1 mW (TE0) and 0.042 mW (TE1), the $E_x$ components of the TE0 (dashed line) and TE1 (dotted line) modes interfere (dash-dot) to form a nodal line 150 nm above the surface. The solid line (right axis) shows the potential energy of a Rb atom above the surface for $\Delta = 1000\Gamma$.**

**(c) Crossing two of the fields shown in (a) leads to a square array of 3D traps 150 nm above the surface, with a spacing of 0.98 $\mu$m.**

**Fig. 2. (Color online) 1D translatable array of 3D traps above a ridge waveguide formed by interfering two modes.**

**Fig. 3. Trap contours (a) in the $x$-$y$ plane and (b) in the $y$-$z$ plane for one of the microtraps ($z^* = 0$) of Fig. 2. For the laser intensities in the text, $\dfrac{\Delta U}{k_B} = 150 \mu\text{K}$ between contours.**

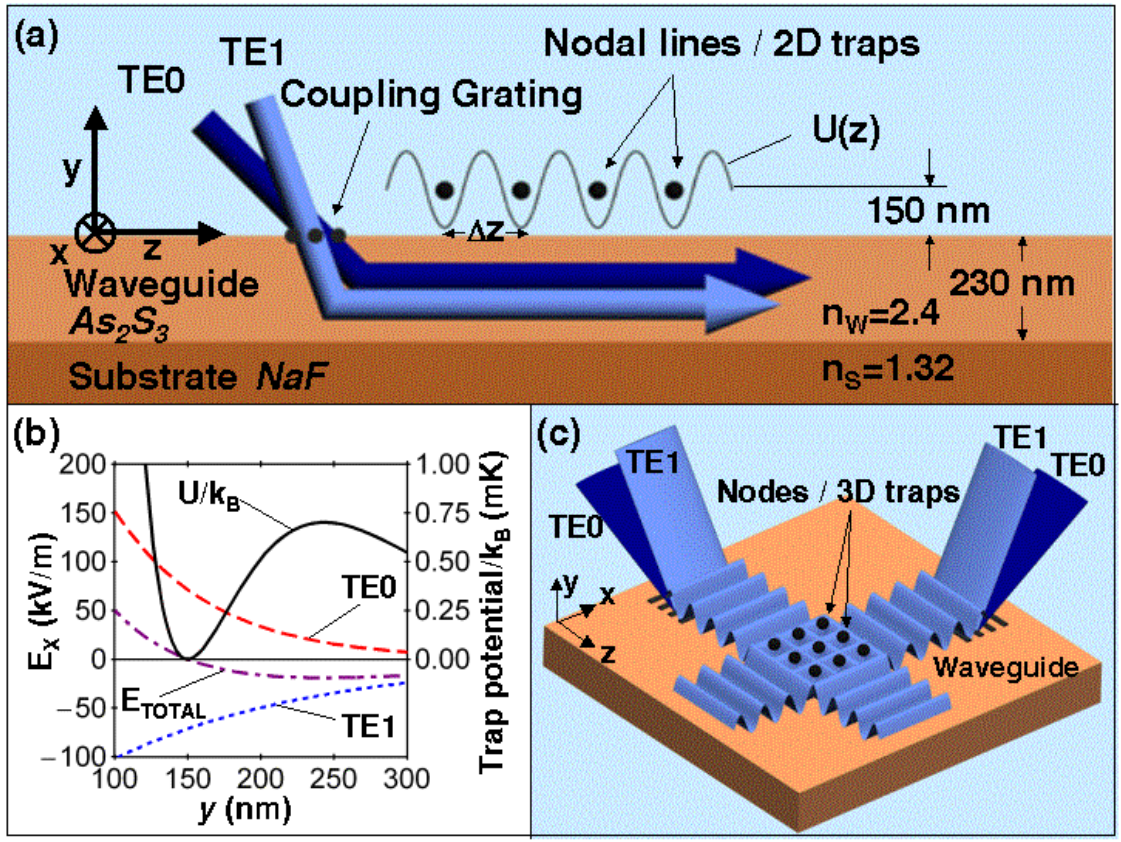

Figure 1

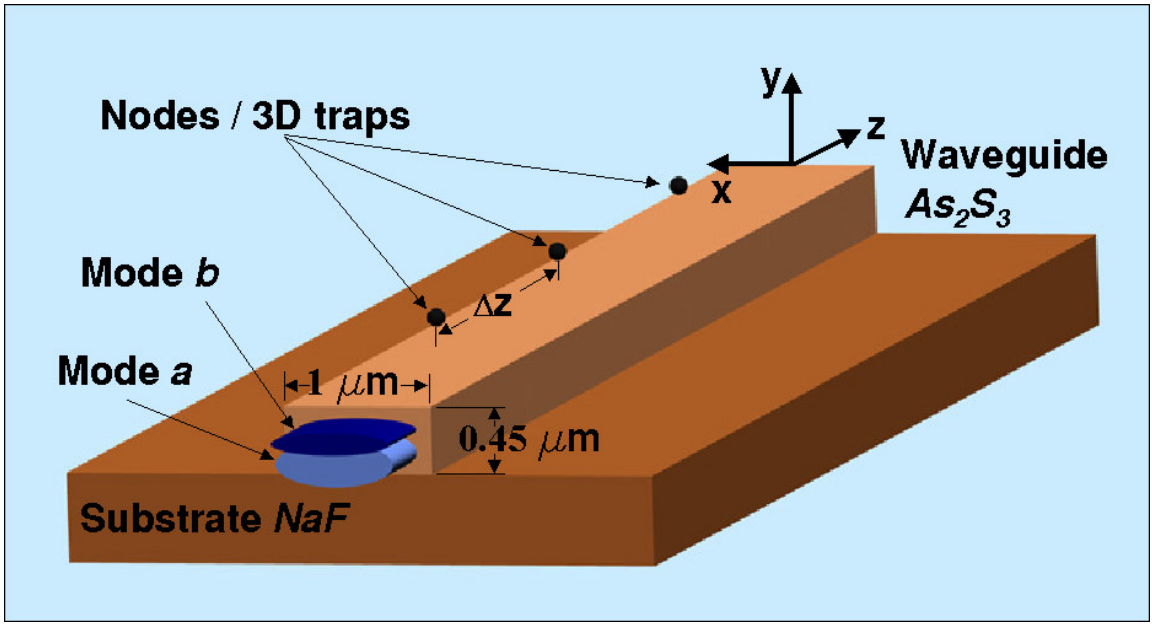

Figure 2

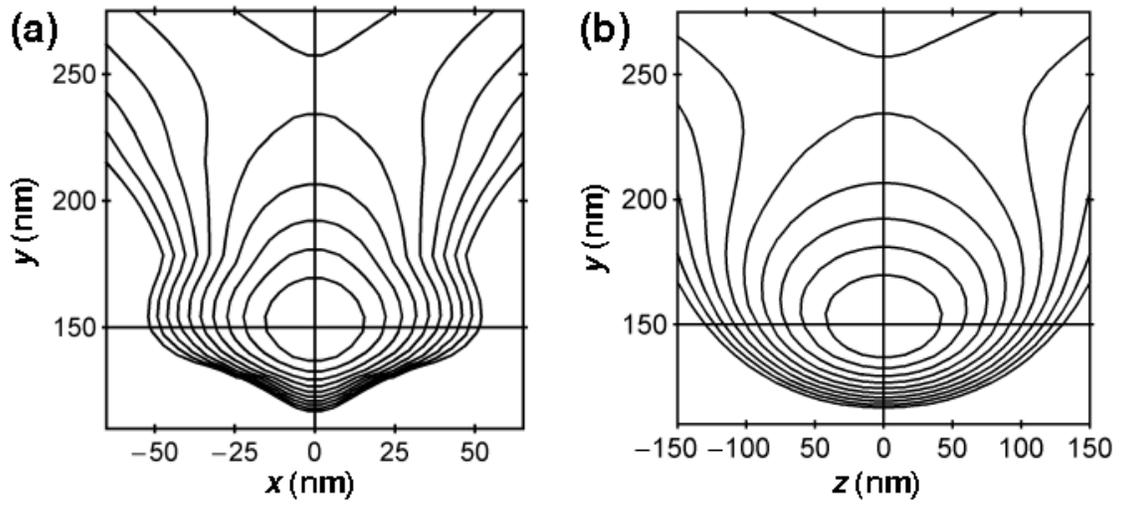

Figure 3

18. We acknowledge helpful conversations with Andrei Modoran and the support of NSF.